\documentclass[useAMS,usenatbib]{mn2e}
\usepackage[figuresright]{rotating}
\usepackage{natbib}
\usepackage[latin1]{inputenc}  
\usepackage{graphicx}
\usepackage{txfonts}
\usepackage{multirow}
\usepackage{epsfig}

\title[Processing of formic acid ice by cosmic ray analogs]{Processing of formic acid-containing ice by heavy and energetic \\ cosmic ray analogs}

\author[A. Bergantini et al.]{A. Bergantini$^{1}$\thanks{E-mail:
alebergantini@gmail.com}, S. Pilling$^{1}$, H. Rothard$^{2}$, P. Boduch $^{2}$, D. P. P. Andrade$^{1}$\\
\\
$^{1}$Universidade do Vale do Paraiba (UNIVAP), Instituto de Pesquisa e Desenvolvimento (IP\&D), Sao Jose dos Campos, SP, Brazil.\\
$^{2}$Centre de Recherche sur les Ions, Matériaux et la Photonique (CEA/CNRS/ENSICAEN Université de Caen-Basse Normandie), CIMAP-CIRIL-Ganil, Caen, France.\\}

\pagerange{\pageref{firstpage}--\pageref{lastpage}} \pubyear{2014}

\begin{document}

\date{Received / Accepted}

\maketitle

\label{firstpage}

\maketitle
\begin{abstract}
Formic acid (HCOOH) has been extensively detected in space environments, including interstellar medium (gas and grains), comets, and meteorites. Such environments are often subjected to the action of ionizing agents, which may cause changes in the molecular structure, thus leading to formation of new species. Formic acid is a possible precursor of prebiotic species, such as Glycine (NH$_2$CH$_2$COOH). This work investigates experimentally the physicochemical effects resulting from interaction of heavy and energetic cosmic ray analogs (46 MeV $^{58}$Ni$^{11+}$) in H$_2$O:HCOOH (1:1) ice, at 15 K, in  ultra-high vacuum regime, using FTIR spectrometry in the mid-infrared region (4000-600 cm$^{-1}$ or 2.5-12.5 $\mu$m). After the bombardment, the sample was slowly heated to room temperature. The results show the dissociation cross section for the formic acid of $2.4 \times 10^{-13}$ cm$^2$, and half-life due to galactic cosmic rays of $\sim8\times10^7$yr. The IR spectra show intense formation of CO and CO$_2$, and small production of more complex species at high fluences.
\end{abstract}


\section{Introduction}
Formic Acid (HCOOH), the simplest carboxylic acid, was first identified in the interstellar medium by \cite{zuck1971}, and subsequently by \cite{winn1975}, using radio telescopes toward the Sagittarius B2 (Sgr B2) region, a large star forming region near the Galactic center. This is one of the few molecules in star forming regions detected both in gas and condensed state (\citealt{schu1999}; \citealt{biss2007}). Formic acid was also detected toward young stellar objects (YSO) by the Infrared Space Observatory (ISO) (\citealt{knez2005}), and its abundance varies from 1 to 5\% with respect to water in solid state (\citealt{biss2007}). In the Solar System, formic acid was detected in comets like C/1996 B2 (\citealt{bock2000}) and C/1995 O1 (\citealt{crov2004}).

Formic acid is a potential precursor of organic and prebiotic molecules such as acetic acid (CH$_3$COOH), methyl formate (HCOOCH$_3$) and glycine (NH$_2$CH$_2$COOH) (\citealt{pill2011}; \citealt{liu2011}). \cite{pill2010} have suggested that the preferable pathway to formation of glycine in astrophysical ices, from carboxylic acids, is due to formic acid.

Observations reveals that the abundance of solid formic acid is a factor of $10^4$ higher than gaseous formic acid in high-mass star-forming regions (\citealt{biss2007}).  \cite{boec2005} show that HCOOH is almost completely destroyed by soft X-rays in gas phase. Therefore, studies in solid state are necessary to better understand the abundances of this molecule in space.

The surface reaction that plays a major role in the production of HCOOH in astrophysical environments is object of discussion. For instance, \cite{benn2011} suggests that formic acid may be formed in interstellar ices, cometary ices, and icy satellites, from reactions involving H$_2$O and CO molecules. The decomposition of H$_2$O forms atomic hydrogen (H) and the hydroxyl radical (OH). The hydrogen atom reacts with CO, producing the formyl radical (HCO). The recombination of (HCO) with the hydroxyl radical (OH) leads to the production of formic acid:
\\
H$_2$O $\rightarrow$ H + OH \\
H + CO $\rightarrow$ HCO	\\
HCO + OH $\rightarrow$ HCOOH

\cite{tiel1982} suggests that the formation of formic acid also starts with H + CO producing HCO, but the pathway to formic acid production is slightly different, and evolves addition of O to produce HCO$_2$ and H to finally HCOOH.

Although it is believed that similar phenomenon occurs in interstellar dust grains, comets and frozen satellites, \cite{iopp2011} present a non-energetic reaction route that reveals to be more efficient to produce formic acid in laboratory conditions:
\\
CO + OH $\rightarrow$ OH-CO\\
OH-CO + H $\rightarrow$ HCOOH

Again we conclude that this molecule needs to be more studied in similar conditions to astrophysical environments.

This study analyzes the results from bombardment of cosmic ray analogs in an astrophysical ice made of a binary mixture of formic acid and water (1:1), and it is an extension of the work in \cite{andr2013}, which did the processing of pure formic acid ice at the same facilities. The main objective of this work is to understand the differences regarding the processing of formic acid ice in the presence of water, a common scenario in several astrophysical environments. The dissociation of molecules in space environments might lead to the production of larger and more complex molecules, including organic and prebiotic material. In this sense, the data acquired experimentally may help in the interpretation of astronomical observations, clarifying the physicochemical processes that take place in some astrophysical sources like molecular clouds, star-forming regions and comets.

\section{Experimental} 

The experiment simulating bombardment of cosmic rays in HCOOH:H$_2$O ice was performed at the IRRSUD beamline of GANIL ("Grand accélérateur National d'ions lourds") in Caen, France, using the CASIMIR set-up of CIMAP-CIRIL ("Centre de recherche sur les Ions, Matériaux et la Photonique").

The chemical changes due to ice irradiation were monitored by Fourier transform infrared (FTIR) spectrometry (Nicolet Magna 550) operating in absorption mode with 2 cm$^{-1}$ resolution, in the mid-infrared region (4000-600 cm$^{-1}$ or 2.5-12.5 $\mu$m).

The samples were purchased commercially, and they were degassed through several freeze-pump-thaw cycles before being mixed in a 1:1 proportion. The gas mixture was deposited in a sample holder with a KBr substrate, previously cooled to 15 K by a helium closed cycle cryostat. The experiment was held in an ultra-high vacuum (UHV) chamber (base pressure lower than $1 \times 10^{-8}$ mbar).

After the sample deposition, an IR spectrum of the pristine ice was collected, and the irradiation of the ice with 46 MeV $^{58}$Ni$^{11+}$ ions began. The ion flux was $2\times10^{9}$ ions cm$^{-2}$ s$^{-1}$, and the integration of the flux over the time gives the final fluence of $1\times 10^{13}$ ions cm$^{-2}$. The ion beam was impinged perpendicularly to the sample. More details regarding the experimental apparatus can be found in \cite{sepe2009}; \cite{pill2010}; and \cite{barr2011}.

\subsection{Ice thickness, mass and deposition rate}
The sample thickness was calculated by (\citealt{pilli2011}):

\begin{equation} \label{thick}
d=\frac{N_0}{6.02\times 10^{23}} \frac{M}{\rho} \times 10^4  \textnormal{[$\mu$m],}
\end{equation}
where $N_0$ is the initial column density (molec cm$^{-2}$), $M$ is the molar mass (g mol$^{-1}$), and $\rho$ is the ice density (1.11g~cm$^{-3}$). The calculated ice thickness was about $\sim$7.4$\mu$m. The penetration depth of 46 MeV Ni ions is much greater than ice thicknesses, and the ions pass through the ice sample with approximately the same velocity (constant cross sections).

The sample mass may be determined by the relation:
\begin{equation} \label{mass}
m = \rho \times \alpha \times d
\end{equation}
where $\rho$ is the density (in g cm$^{-3}$), $\alpha$ is the surface area of the KBr crystal that holds the sample ($\alpha$ $\thicksim$ 0.55 cm$^2$), and $d$ is the ice thickness (in cm).
The calculated sample mass at the beginning of the experiment was $m=453$~$\mu g$. No deposition of material was made after the beginning of the irradiation.


\section{Results and discussion}

\subsection{Identification of the IR bands}

Table~\ref{tab:pos} shows the band positions of the formic acid in this work compared to the literature. The symbols $\nu_{S}$ and $\nu_{B}$ indicates the stretching and angular deformation modes, respectively. The band positions in this work are more similar to those of \cite{biss2007} in comparison with the work of \cite{cyri2005} and \cite{andr2013}, probably because of the presence of water in the ice.

\begin{table}
\caption{Comparison of the positions of the main formic acid vibrational bands in this work with literature.}
\label{tab:pos}
  \setlength{\tabcolsep} {2pt} 
\begin{tabular}{l c c c c}
\hline
   & This work              & Bisschop et al.  & Cyricac \&     & Andrade et al.\\
   &                        & (2007)           & Pradeep (2005) & (2013) \\
\hline
  Mode  & \multicolumn{4}{c}{Position (cm$^{-1}$)} \\
  \hline
$\nu_{S}$(O-H) & 3240 & 3280 & 3115 & 3109  \\
$\nu_{S}$(C-H)  & 2932 & 2953 & 2954 & 2954  \\
$\nu_{S}$(C=O) & 1685 & 1650 & 1708 & 1709  \\
$\nu_{S}$(OH/CH)& 1380 & 1387 & 1389 & 1390 \\
$\nu_{S}$(C-O) & 1211 & 1211 & 1373 & 1373 \\
$\nu_{B}$(C-H) & 1067 & 1073 & 1072 & 1072 \\
\hline
Ice: & H$_2$O:HCOOH & \multicolumn{3}{c}{HCOOH pure} \\
\hline
\end{tabular}
\end{table}

\begin{figure}
\centering
\resizebox{\hsize}{!}{\includegraphics{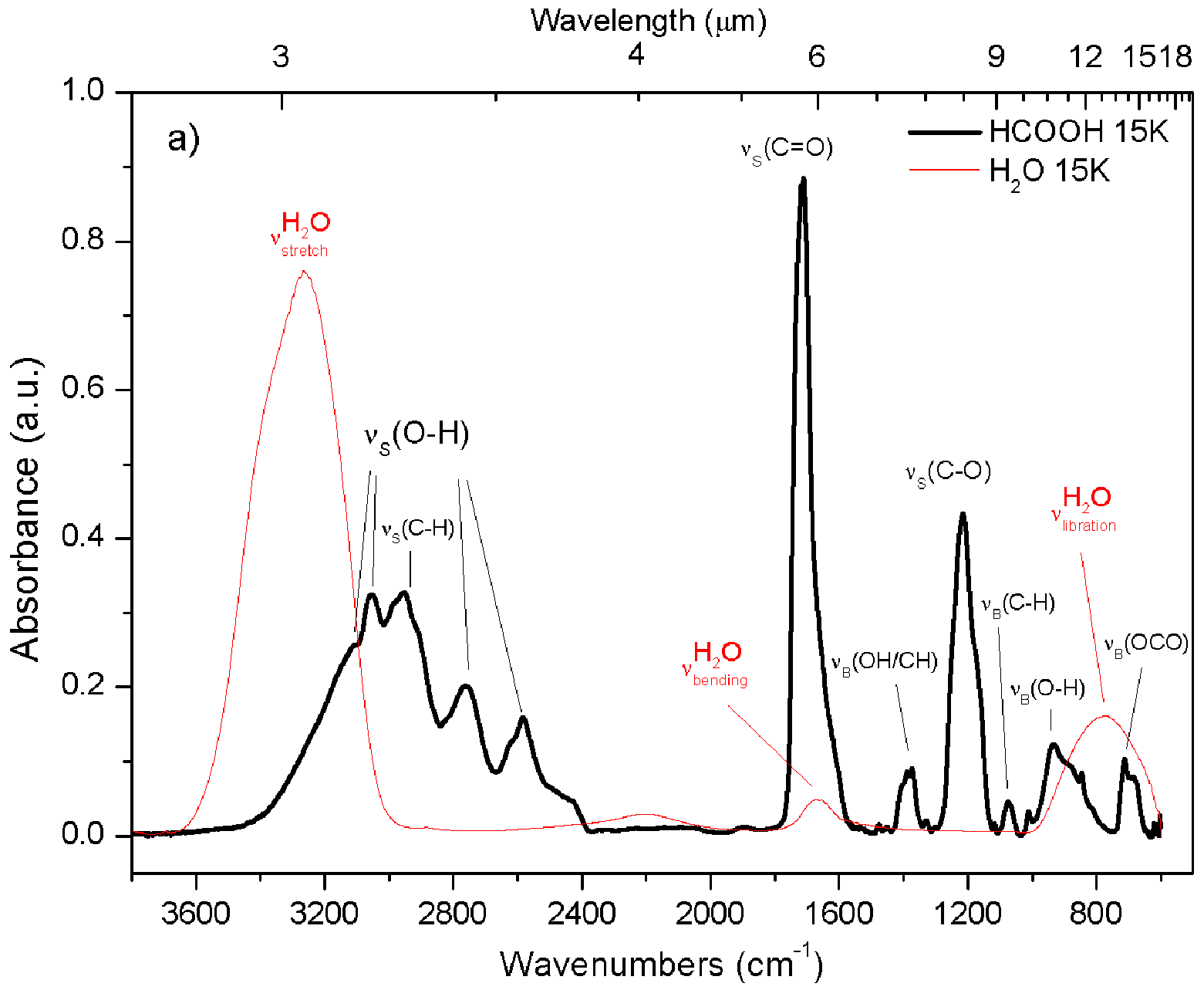}}
\caption{Vibrational spectra of pure formic acid ice (black thick) and pure water ice (red thin) at 15K. Notice the partial overlap of some bands of these two molecules (between $\thicksim$3300 and $\thicksim$3000 cm$^{-1}$, between $\thicksim$1800 and $\thicksim$1600 cm$^{-1}$, and between 900 and 700 cm$^{-1}$) which led to the choice of the $\nu_{S}$(C-O) band, in 1211 cm$^{-1}$ as reference band for the formic acid molecule. These data were acquired by Bergantini et al. (\textit{in preparation}).}
\label{fig:comp_agua_acid_puro}
\end{figure}

To better understand the influence of the water in the spectrum, Figure~\ref{fig:comp_agua_acid_puro} shows spectra of pure formic acid (black thick) and pure water (red thin), both at 15 K. The stretching mode of the (C=O) band in $\thicksim$1685 cm$^{-1}$ is the most characteristic spectral signature of the formic acid (\citealt{biss2007}), but it suffers partial overlap with the $\nu_{B}$(OH) band of the solid water in $\thicksim$1660 cm$^{-1}$, so it will not be the reference for the formic acid molecule in this work.

Another relevant band to characterize formic acid is the $\nu_{S}$(C-O) band, in 1211 cm$^{-1}$. Even though this is not the strongest band of the formic acid, it has been detected toward high mass star forming regions (\citealt{schu1999}) and in molecular clouds (\citealt{biss2007}), and it is not influenced by the water bands. Therefore, the $\nu_{S}$(C-O) band, in 1211 cm$^{-1}$, will be the reference for formic acid in this study.

\begin{figure}
\centering
\resizebox{\hsize}{!}{\includegraphics{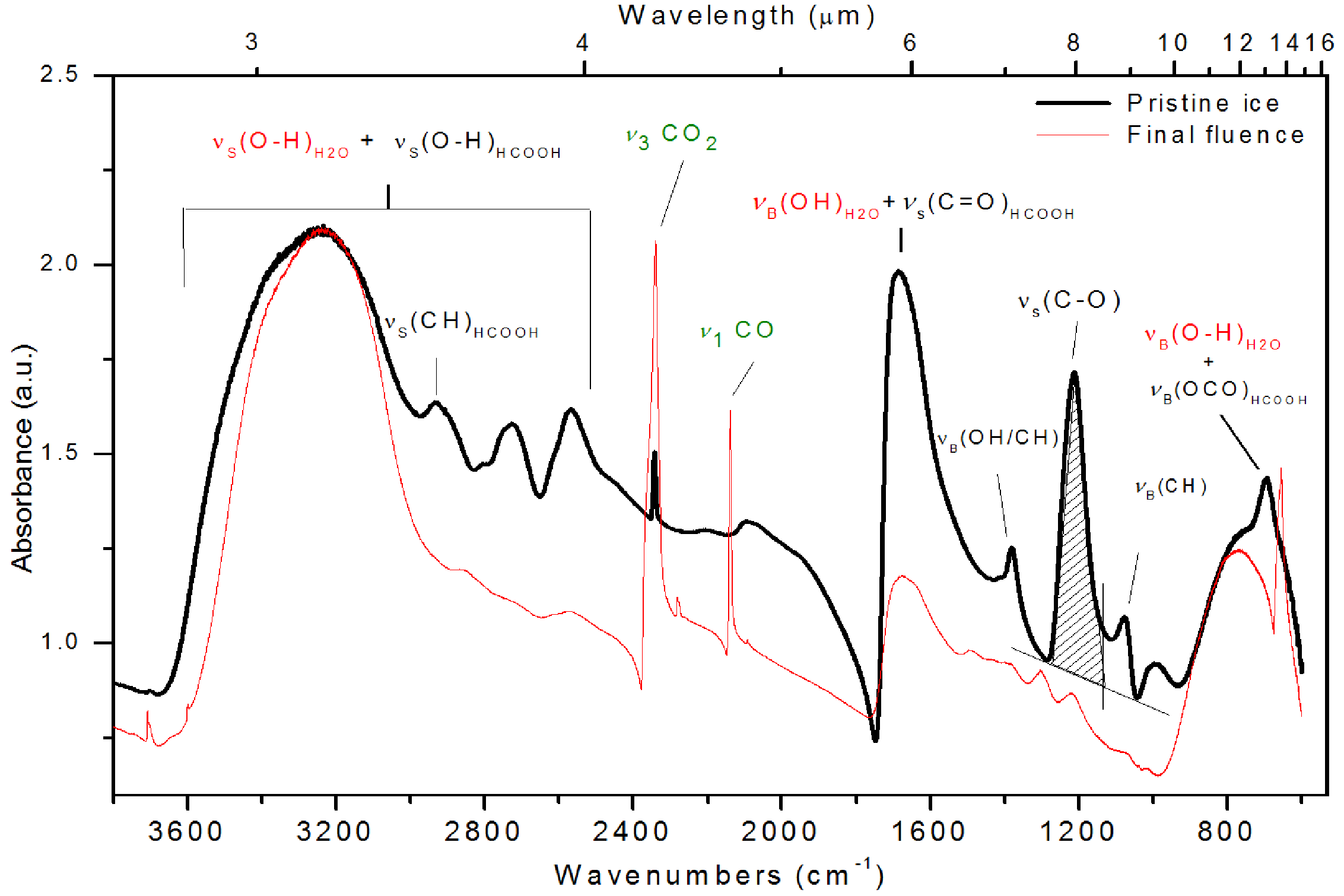}}
\caption{Vibrational spectrum of the H$_2$O:HCOOH ice before (black thick) and after the bombardment (red thin) by heavy ions at the final fluence of 10$^{13}$ ions/cm$^{-2}$ (red). The $\nu_{S}$(C-O) band was used to characterize the HCOOH molecule.}
\label{fig:spec_virg_final_fluence}
\end{figure}

Figure~\ref{fig:spec_virg_final_fluence} shows the infrared spectra of the pristine (black thick) and irradiated (red thin) HCOOH:H$_2$O ice. The large band between $\sim$3600 and $\sim$3000 cm$^{-1}$ is heavily influenced by the stretching mode of water, and it remains virtually unchanged after the irradiation, probably because of constant water layering.

Some important vibrational modes of the formic acid, such as $\nu_{S}$(C=O) and $\nu_{S}$(C-O), had their intensity severely diminished due to the bombardment. On the other hand, the 2145 cm$^{-1}$ and 2350 cm$^{-1}$ bands, which refer to carbon monoxide (CO) and carbon dioxide (CO$_2$) respectively, had their intensity increased and are the main produced species due to the ice processing.

\begin{figure}
\centering
\resizebox{\hsize}{!}{\includegraphics{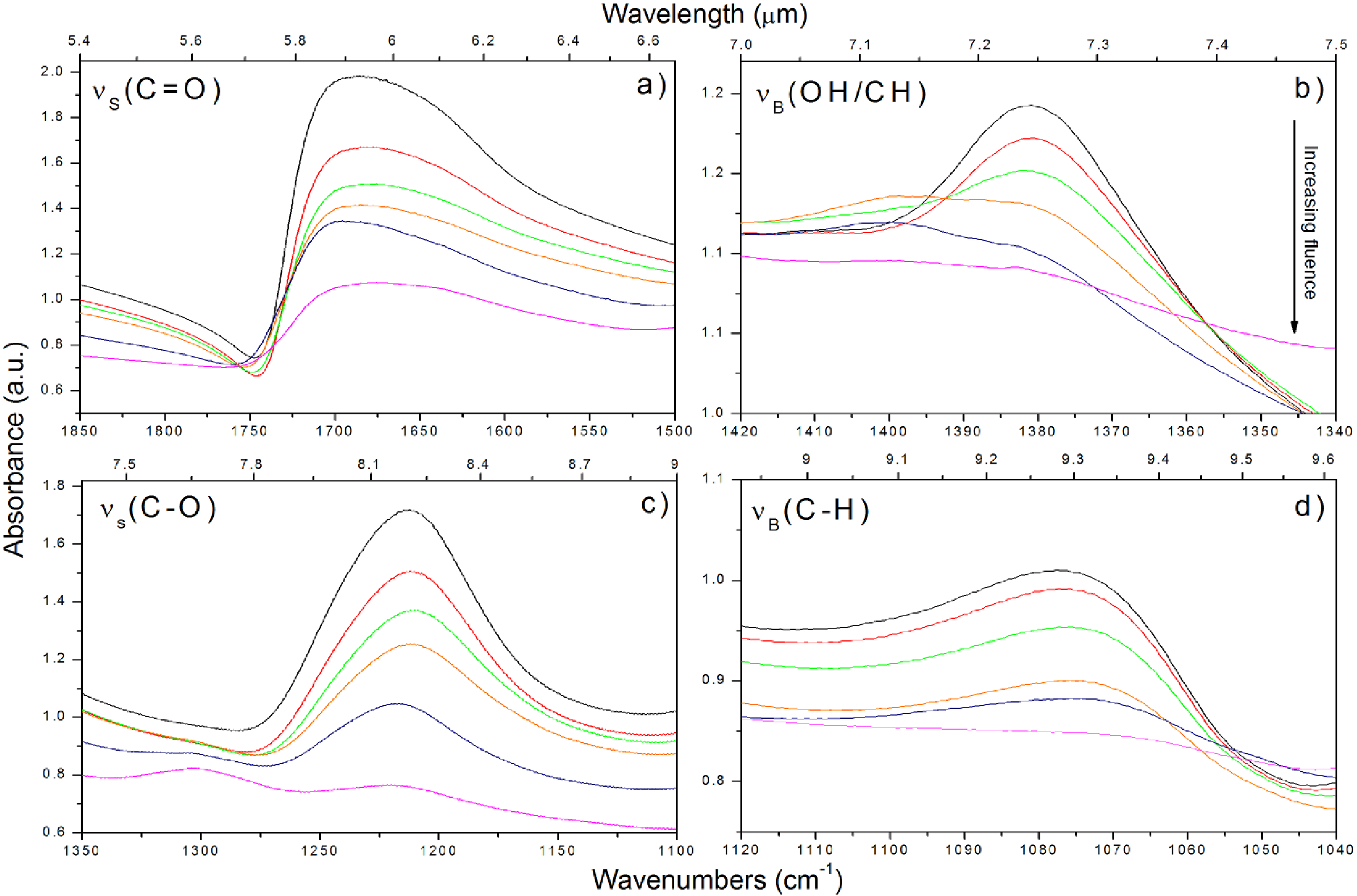}}
\caption{Evolution of the areas for selected vibrational IR bands of formic acid under heavy bombardment. The fluence increases from top to bottom in all the four Figures. The top axis indicates the wavelength in micrometers. The asymmetric profile of the bands is probably related with the water influence.}
\label{fig:quatr_graf}
\end{figure}

Figure~\ref{fig:quatr_graf} shows the variation in the area of the main formic acid bands in function of the fluence. Fig.~\ref{fig:quatr_graf}a) shows the $\nu_{S}$(C=O) band in 1685 cm$^{-1}$; Fig.~\ref{fig:quatr_graf}b) $\nu_{S}$(OH/CH) band in 1380 cm$^{-1}$; Fig.~\ref{fig:quatr_graf}c) $\nu_{S}$(C-O) band in 1211 cm$^{-1}$; and Fig.~\ref{fig:quatr_graf}d) $\nu_{S}$(C-H) band in 1076 cm$^{-1}$. The asymmetric profile of the bands shown in Figures \ref{fig:quatr_graf}a), \ref{fig:quatr_graf}b), and \ref{fig:quatr_graf}d) is probably related with the water influence, since it is not observed in the C-O bond.

\subsection{Column density}
The rewritten Lambert-Beer equation was used in order to calculate the column density ($N$) of the formic acid (\citealt{barr2011}):
\begin{equation}
\label{dcolumn}
N = ln(10)\frac{\int_{\nu i}^{\nu f} A'_{(\nu)}d\nu}{A}
\end{equation}
where $N$ is the column density, $A'_{(\nu)}$ is the absorbance coefficient at given frequency ($\nu$), and $A$ is the band strength (or \textit{A-value}. The \textit{A-values} from literature are shown in Table~\ref{tab:big}), and $\nu{i} - \nu{f}$ is the integration range. The results show, based on the $\nu_{S}$(C-O) band, that the initial column density ($N_0$) of formic acid was $\sim8\times 10^{18}$ molec/ cm$^{2}$.

The variation in the column density over the fluence allows us to quantify the destruction of the parent molecule and the formation of daughter molecules. The initial (N$_i$) and final (N$_f$) column density (calculated by Equation~(\ref{dcolumn})) are shown in Table~\ref{tab:big}. The peak position of the main bands of the ice, the integration range used for each band and the values of the band strengths from literature (or \textit{A-value}) are also given.

\begin{table*}
\centering
\caption{Position of the detected bands, with the correspondent assignments, the integration range used to obtain the areas under the bands, the band strengths from literature and the initial and final calculated column density.}
\label{tab:big}
\begin{tabular} { c c c c c c c l }
\hline
Band position   & Molecule      & Vibrational       & Integration          & N$_i$                 & N$_f$                   & Band strength         & Reference of the\\
(cm$^{-1}$)     &               & mode              & range (cm$^{-1})$    & (molec cm $^{-2}$)    & (molec cm $^{-2}$)      & (cm molec$^{-1}$)     & Band strength  \\
\hline
2340            & CO$_2$        & $\nu_3$ stretch   & 2382 - 2308          & $4.3\times 10^{16}$  & $5.5\times 10^{17}$    & $7.6\times 10^{-17}$ & \citet{gera1995}  \\
2145            & CO            & $\nu_1$ stretch   & 2147 - 2120          & traces               & $8.2\times 10^{17}$    & $1.1\times 10^{-17}$ & \citet{gera1995}  \\
1685            & HCOOH         & $\nu_S$(C=O)      & 1746 - 1430          & $5.8\times 10^{18}$  & $1.4\times 10^{18}$    & $6.7\times 10^{-17}$ & \citet{schu1999}  \\
1380            & HCOOH         & $\nu_B$(OH/CH)    & 1398 - 1357          & $4.8\times 10^{18}$  & $6.1\times 10^{17}$    & $2.6\times 10^{-18}$ & \citet{schu1999}  \\
1211            & HCOOH         & $\nu_S$(C-O)      & 1273 - 1151          & $8.0\times 10^{18}$  & $4.8\times 10^{17}$    & $1.5\times 10^{-17}$ & \citet{huds1999}  \\
\hline
\end{tabular}
\end{table*}

This table shows that the final column density of the $\nu_S$(C=O) formic acid band is approximately one order of magnitude higher in comparison with the other formic acid bands. This is most certainly due to the influence of the $\nu_B$(O-H) water band in the same region of the spectrum. Water is one of the main products obtained by Andrade et al. (2013) due to irradiation of pure formic acid by heavy and energetic ions. Therefore, and also because this is a 1:1 H$_2$O:HCOOH mixture, the overlap with the bending mode of water is not a negligible effect in the $\nu_S$(C=O) band.

\begin{figure}
\centering
\resizebox{\hsize}{!}{\includegraphics{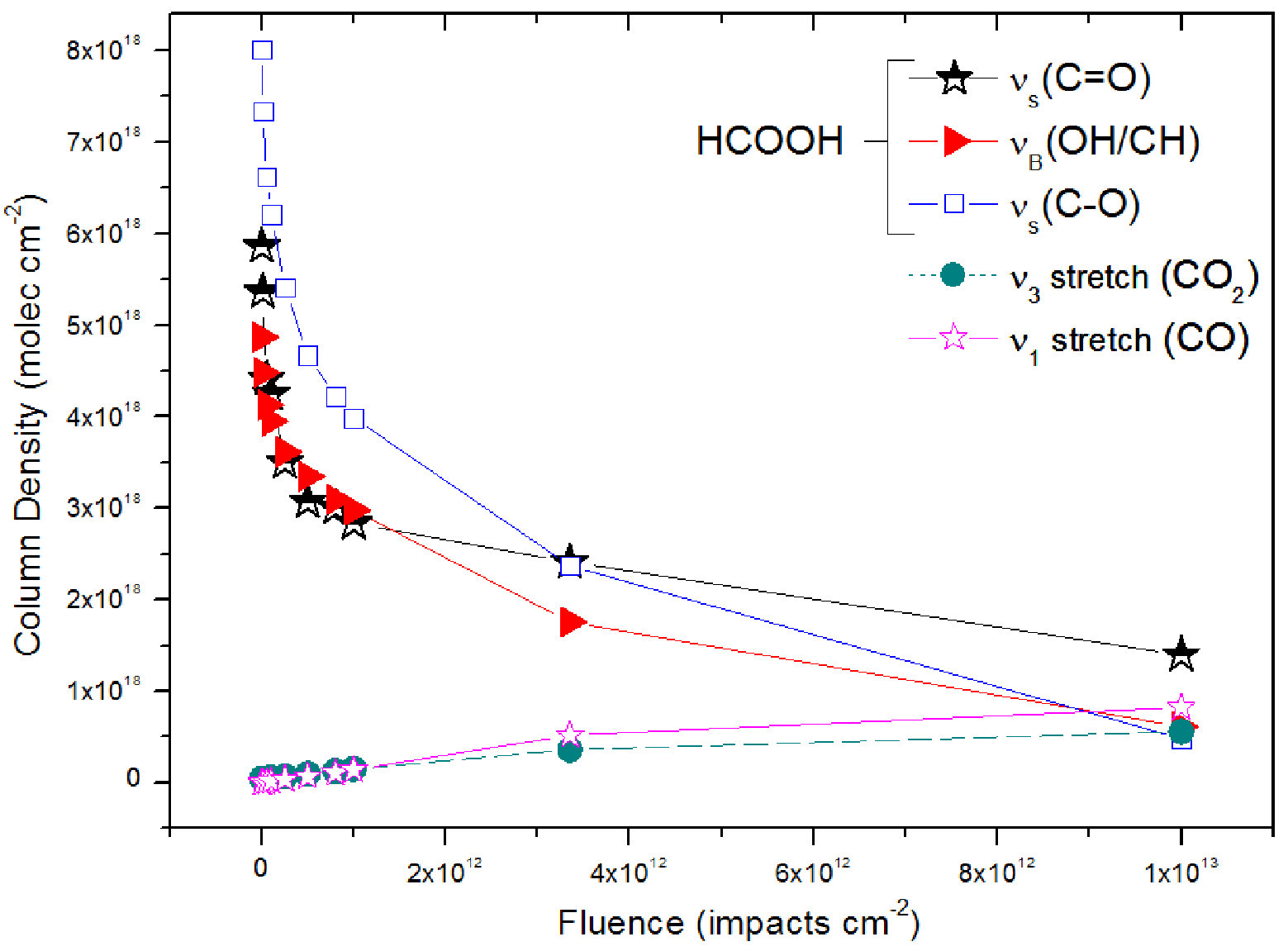}}
\caption{Variation of the column density for the main IR bands of the ice, for both parent and daughter molecules, calculated over the fluence. The points were obtained experimentally and the lines are just for guidance. The two bottom lines refers to carbon dioxide and carbon monoxide.}
\label{fig:new_column_densy}
\end{figure}

During the irradiation, a series of spectra were collected from time to time in order to monitor the changes in the ice due to the irradiation. By measuring the variation in the area of the bands over the fluence, and assuming that the band strength does not change significantly, it is possible to measure how the column density varies over the fluence both for parent and daughter species. Figure~\ref{fig:new_column_densy} shows the the column density for the points collected experimentally. The column density of the formic acid drops by half at fluences near $1\times 10^{12}$ impacts cm$^{-2}$. In fluences around $7.6\times 10^{17}$, the abundance of the daughter species are roughly the same as the parent molecule (using the C-O band as reference for formic acid).

\begin{figure}
\centering
\resizebox{\hsize}{!}{\includegraphics{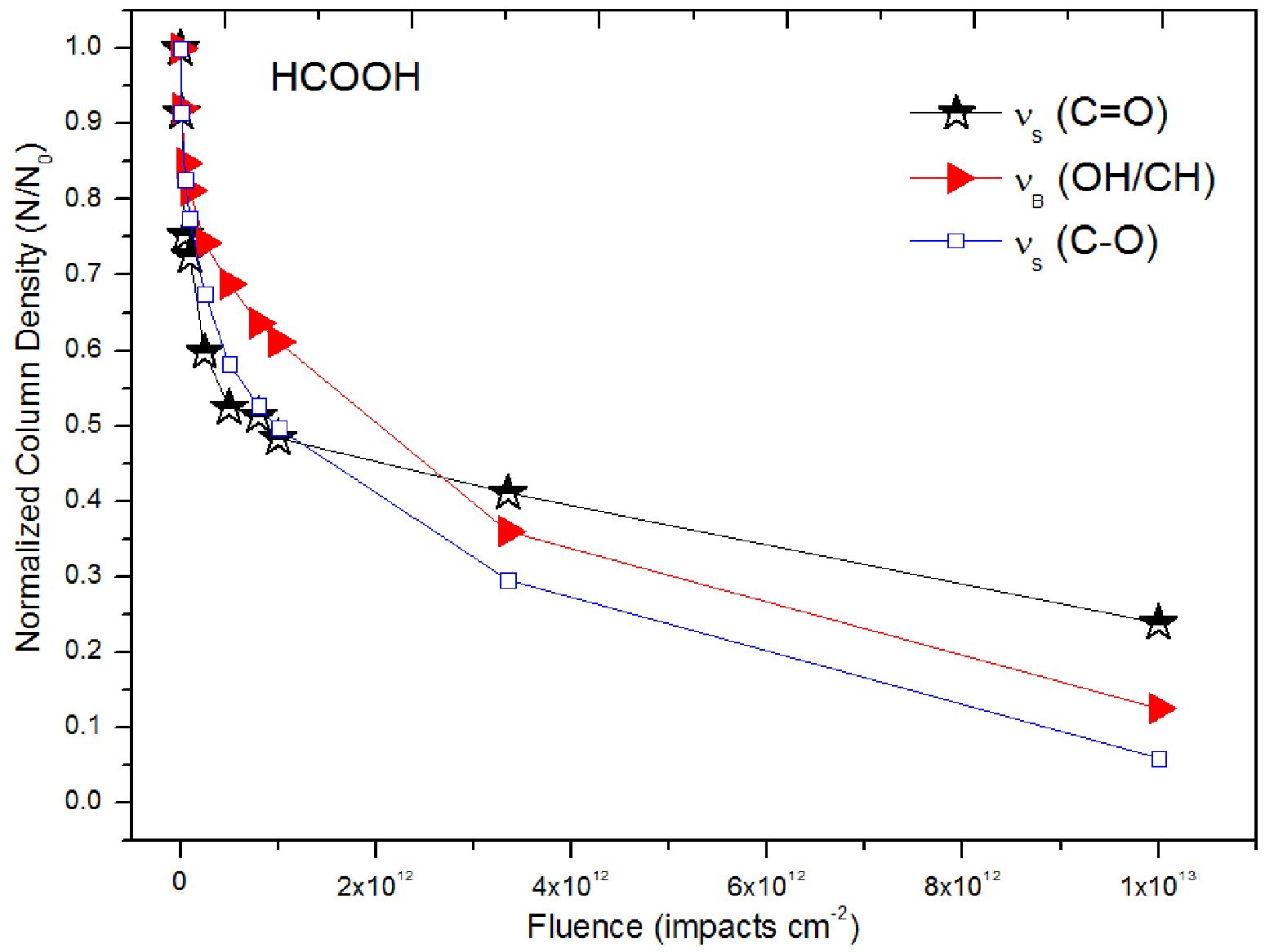}}
\caption{Normalized column density over the fluence for the main formic acid IR bands. The lines are just for guidance.}
\label{fig:norm_new_column_densy}
\end{figure}

As seen in Fig.~\ref{fig:new_column_densy} and in Table~\ref{tab:big}, the values for the initial column density for the formic acid bands are not the same, as they should be theoretically. These differences are related mainly with differences regarding the values of the band strengths from literature and the actual band strengths may vary from one ice to another. Because of this, Figure~\ref{fig:norm_new_column_densy} shows the normalized column density for formic acid bands. This normalization was made by the quotient of the column density at given fluence by the initial column density (N/N$_0$).

As discussed by \cite{barr2011}, the \emph{carbon budget} is a way to test if the band strength values from literature are compatible with the experiment. Using the C-O bond as reference, the variation in the formic acid column density indicates that $\thicksim7\times 10^{18}$ molecules were destroyed per cm$^2$. The sum of the produced (daughter) molecules may not be higher than this figure. The results indicate that the combination of the main produced molecules, CO and CO$_2$, is $\thicksim1\times 10^{18}$ molecules per cm$^2$, which is consistent with the amount of parent molecules destroyed.

Table~\ref{tab:carbon_budget} shows a summary of the variation of the column densities (N$_f$ - N$_0$), as well as the absolute yield (number of molecules destroyed or produced per impact).

\begin{table}
\caption{Variation in the column density for the parent and daughter molecules and absolute yield (molecules destroyed of formed per impact).}
\label{tab:carbon_budget}
   \setlength{\tabcolsep} {2pt} 
\begin{tabular} { c c c c }
\hline
Molecule   & Wavenumber & Variation in the column  & Absolute yield        \\
           & cm$^{-1}$  & density (N$_f$ - N$_0$)  & (molec impact$^{-1}$) \\
\hline
HCOOH      & 1211       & $-7\times 10^{18}$       & $7\times 10^{5}$    \\
CO         & 2145       & $8\times 10^{17}$        & $8\times 10^{4}$    \\
CO$_2$     & 2340       & $5\times 10^{17}$        & $5\times 10^{4}$    \\
\hline
\end{tabular}
\end{table}

The produced-destroyed molecules ratio is approximately 0.2. This is mainly due to sputtering and to the production of molecules that do not have enough yield to be detected in the IR spectrum. Some of these new molecules can be detected at higher temperatures, when other constituents of ice, such as water, evaporate. Molecules observed during the sample heating are shown in Table~\ref{tab:newbands}, in Section 3.4.


\subsection{Determination of the cross sections}

The decreasing in the column density of the formic acid bands (see Fig.~\ref{fig:new_column_densy}) is caused mainly by the sputtering induced by heavy ions and the destruction of the molecule, with subsequent formation of new species (\citealt{sepe2009}). As discussed by \cite{barr2011}, \cite{pill2012}, and \cite{andr2013}, the relation between the variation in the column density over the fluence ($F$) can be described by:
\begin{equation}
\label{sigma_d}
N = N_0~exp(-\sigma_d~F)
\end{equation}

The formation cross section calculated for the main bands of the formic acid are shown in Table~\ref{tab:sigmad}, which also compares the results with those reported by \cite{andr2013}, in a similar experiment.

The formation cross section for the daughter molecules, produced directly from the parent, can be calculated by:

\begin{equation}\label{eq:daughter1}
\frac{N_k(F)}{N_\nu,0} = \sigma_{f,k} \big( F - \frac{\sigma_{d,\nu}+\sigma_{d,k}}{2}F^2 \big)
\end{equation}
where $N_k$(F) is de column density of the daughter molecule in a given fluence; $N_\nu,0$ is the parent column density at $F=0$; $\sigma_{f,k}$ is the formation cross section for the daughter molecule k; $\sigma_{d,\nu}$ is the destruction cross section of the parent molecule; and $\sigma_{d,k}$ is the destruction cross section for the daughter molecule.

For low fluences, we may assume that the influence of the $\sigma_{d,\nu}$ factor is negligible, (for details, please refer to \citealt{andr2013} and \citealt{barr2011}), and equation \ref{eq:daughter1} can be easily fitted to provide the values of the formation and destruction cross sections for the daughter molecules. These results are also shown in Table~\ref{tab:sigmad}.

\begin{table*}
\caption{The calculated values of the destruction cross sections ($\sigma_d$) for three formic acid bands, as well as the values of the formation cross section $\sigma_f$ for the two main produced species, CO and CO2, and compares the results with literature.}
\label{tab:sigmad}
\begin{tabular}{ c c c c c }
\hline
  Molecule & Vibrational         & Position                   & \multicolumn{2}{c}{Destruction Cross Section ($\sigma_d$) (cm$^2$)}\\

           & Mode                & (cm$^{-1}$)                &  This work            &  Andrade et al. 2013         \\
\hline
HCOOH      & $\nu_{S}$(C=O)         & 1685                    & $8.6\times 10^{-14}$  &  $1.4\times 10^{-13}$   \\
HCOOH      & $\nu_{S}$(OH/CH)       & 1380                    & $1.8\times 10^{-13}$  &  -                      \\
HCOOH      & $\nu_{S}$(C-O)         & 1211                    & $2.4\times 10^{-13}$  &  $1.3\times 10^{-13}$   \\
\hline
           &                        &                         & \multicolumn{2}{c}{Formation Cross Section ($\sigma_f$) (cm$^2$)}\\
           &                        &                         & This work             &  Andrade et al. 2013         \\
CO         & $\nu_1$ stretch        & 2145                    & $6.3\times 10^{-14}$  &  $3\times 10^{-14}$    \\
CO$_2$     & $\nu_3$ stretch        & 2340                    & $6.7\times 10^{-14}$  &  $3\times 10^{-14}$    \\
\hline
           \multicolumn{3}{c}{Ice composition:}            & H$_2$O:HCOOH          & HCOOH pure \\
\hline
\end{tabular}
\end{table*}


\subsection{Sample heating}

\begin{figure}
\centering
\resizebox{\hsize}{!}{\includegraphics{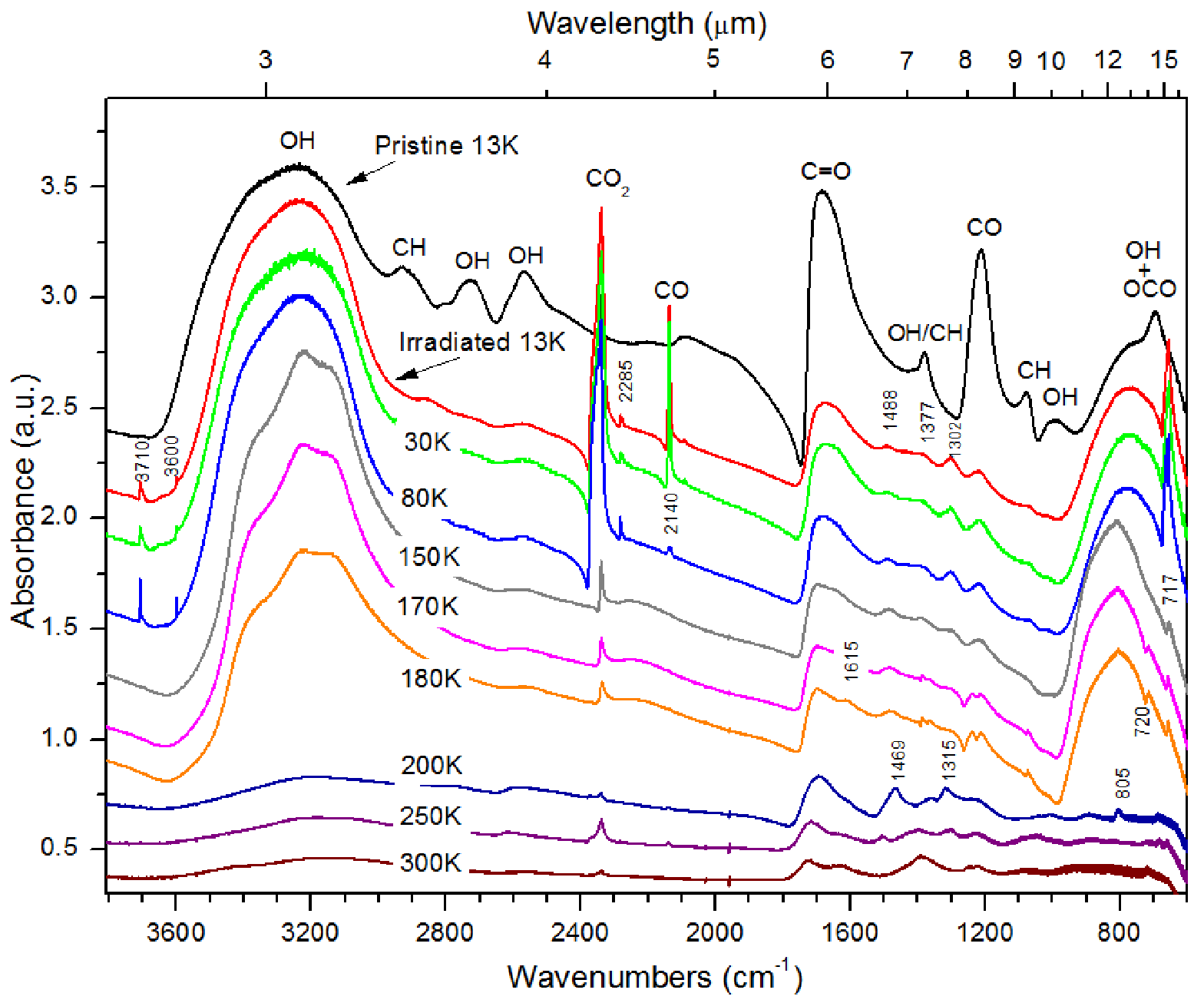}}
\caption{Spectra collected after the bombardment, in different temperatures, from 13 K to 300 K. The observed new bands are indicated by the correspondent wavenumber. Most of the molecules desorbs together with the water, in temperatures above 180 K. Spectra have been offset for comparison purposes.}
\label{fig:heating}
\end{figure}

After the bombardment, the ice sample was slowly heated (ramp of 1 K min$^{-1}$) to room temperature. New spectra were collected during the heating in order to analyze the possible presence of species not detected in the IR spectrum at low temperatures. The heating spectra is shown in Figure~\ref{fig:heating}.

In the 3710-3600 cm$^{-1}$ region, no peaks are visible in pristine ice, but the processed ice has two visible peaks. The wide water band around 3200 cm$^{-2}$ starts to change its profile in temperatures above 150 K, probably because of the structure change from amorphous to crystalline. This band evaporates in temperatures above 180 K. The C-H and O-H bands from formic acid and water in the region between 2850 and 2200 cm$^{-1}$ are not distinguishable in the heating spectra.

The $^{12}$CO$_2$, $^{13}$CO$_2$, and CO bands, at 2340cm$^{-1}$, 2285cm$^{-1}$, and 2145cm$^{-1}$ respectively, have their intensity highly reduced in temperatures above 80 K. Only the $^{12}$CO$_2$ is noticeable after 150 K, this is probably due to a small contamination.

Some IR bands are visible only after the sample heating. We highlight the following: 1488, 1377, and 1302 cm$^{-1}$ at 30 K; 717 and 720 cm$^{-1}$ at 150 K; 1615 cm$^{-1}$ at 180 K; 1469, 1315, and 805 cm$^{-1}$ in temperatures above 200 K.

The band at 1302 cm$^{-1}$, attributed C-H bond, is not visible in the pristine ice, but is clearly distinguishable in the irradiated ice and i only evaporates in temperatures higher than 150 K. The peak at 1125 cm$^{-1}$, assigned to the stretching mode of C-O, transforms in to two weak peaks in the 170 K and 180 K spectra. This could be explained by a structure change, from amorphous to crystalline ice. This band disappears together with water in temperatures above 180 K. The very weak band at 1615 cm$^{-1}$, in the 170 and 180 K spectra, is tentatively assigned to the C=C bond, which is not present in the virgin ice, and is thus due to a new specie.

The band at 1469 cm$^{-1}$ becomes more strong at 200 K. It is assigned to C-H bond, an it is probably due to the formation of a less volatile specie, which probably suffers any phase transition in this temperature, just like the 1315 cm$^{-1}$ bond, which is stronger in the 200 and 250 K spectra.

The initially very wide band around 700 cm-1, due to water and the OCO bond, disappears in temperatures above 180 K and turns in to a very weak band (805 cm$^{-1}$) in the 200 K spectrum, assigned to a C-H bond of a less volatile specie.

The relatively strong CO$_2$ (2340 cm$^{-1}$) band decreases significantly in temperatures higher than 80 K, but traces of this molecule are detected in all temperatures. This could be explained both due to small contamination or the formation of a new specie.

Table~\ref{tab:newbands} shows some suggestions about new bands detected in the spectrum during the sample heating, according to IR spectroscopic data available from \citet{naka1977}.

\begin{table*}
\centering
\caption{Assignment of selected bands in the IR spectra after bombarded and heating from 15 K to 300 K. The suggested assignments were selected from \citet{naka1977}.}
\label{tab:newbands}
\begin{tabular} {c l c l l}
\hline
Pos.(cm$^{-1)}$ & Temp.             & Tentative assignment      & Mode                     & Functional group \\
\hline
3710            & 13 K              & O-H                       & stretch                       & carb. acid, alcohol       \\
2285            & 13 K              & C$\equiv$C                & stretch                       & alkyne       \\
1302            & 13 K              & C-H                       & sciss., bending           & alkane      \\
1125            & 150 K             & C-O                       & stretch                       & carb. acid, alcohol \\
717             & 150 K             & C-H                       & bend                          & alkene \\
1615            & 170 K             & C=C                       & stretch                       & alkyne    \\
1469            & 200 K             & C-H                       & sciss., bending           & alkane \\
1315            & 200 K             & ?                         &                               &  \\
805             & 200 K             & C-H                       & bend                          & alkene \\
\hline
\end{tabular}
\end{table*}


\section{Astrophysical implications}

Formic acid is found in several astrophysical environments such as star forming regions, dense molecular clouds, comets, and small bodies in the Solar System (see for example \citealt{biss2007}; \citealt{crov1998}; \citealt{dish1995}). \cite{cern2012} detected formic acid and the methoxy radical (CH$_3$O), among other molecules, toward the cold and dense core B1, and attributes the presence of these species to non-thermal desorption of the surface of dust grains due to cosmic rays and secondary photons. \cite{brou2013} highlights the role of the HCOOH molecule in the production of dimethyl ether (CH$_3$OCH$_3$) and the methyl formate (HCOOCH$_3$) (gas phase) toward \textit{Orion-KL}, a nearby star-forming region. \cite{oshi2012} raises the possibility of formation of formic acid in Mars due to irradiation of $^{60}$CO in CO$_2$ hydrate.

\subsection{Dependence of the destruction cross section on electronic stopping power}

In astrophysical environments, interaction between energetic particles and matter (interstellar dust grains, frozen surfaces of comets, and icy satellites, etc) is a common event. In the ion-solid interaction, the projectile transfers energy continuously to the target atoms, through successive collisions. The rate of kinetic energy loss as a function of penetration of the ion into solid is called electronic Stopping Power (for fast and highly energetic ions). This interaction generates heating, causes changes in the target structure, and produces chemical reactions and sputtering.

As discussed by \citealt{brow1984}; \cite{sepe2010};  \cite{barr2011}, the destruction cross section of heavy ions may follow a power law as function of the electronic stopping power of the target:
\begin{equation}
\label{eq:sigmaag}
\sigma_d \approx~a~S_{e}^{n}
\end{equation}
were S$_e$ is the electronic stopping power, $\emph{a}$ and $\emph{n}$ are constants.

Previous studies suggest that  $n=1.5$ is a good approximation for this molecule (\citealt{andr2013}). The electronic stopping power (S$_{e}$) in function of the ion energy (from 10 KeV to 10 GeV) is calculated by the software \textsc{srim} (\textit{The Stopping and Range of Ions in Matter} - \citealt{zieg1985}), and the $\emph{a}$-value was determined using the $\sigma_d$ value experimentally obtained (see Table~\ref{tab:sigmad}). This results in $a = 1.75 \times 10^{-17}$ ($10^{15}$$~$molec$~$eV$^{-1}$).

\begin{figure}
\centering
\resizebox{\hsize}{!}{\includegraphics{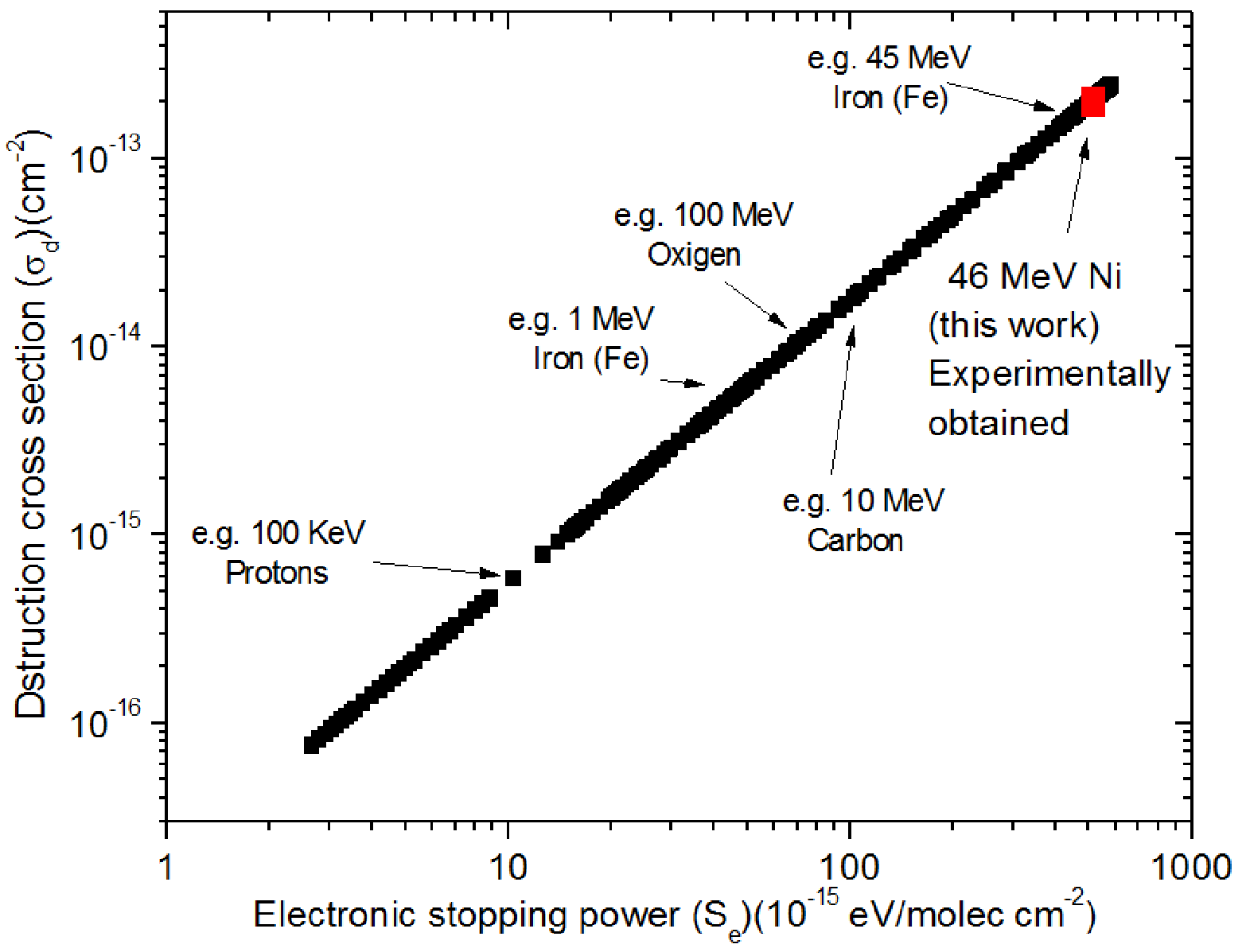}}
\caption{Dependence in the destruction cross section ($\sigma_d$) for HCOOH as function of the electronic stopping power of fast ions. The red square corresponds to the point experimentally obtained with 46 MeV $^{58}$Ni$^{11+}$ projectile. The other points were calculated by Eq. 6.}
\label{fig:sigma_d}
\end{figure}

The fitting of Eq.~\ref{eq:sigmaag} over the electronic stopping power gives the destruction cross section for any cosmic ray ion. This result is shown in Figure~\ref{fig:sigma_d}.

\subsection{Half-life of HCOOH due to cosmic rays}
From the calculated destruction cross section and the estimated flux of galactic cosmic rays ($\Phi_{GCR}$), it is possible to determine the typical half-life ($\tau_{1/2}$) of the formic acid ice in interstellar medium due to heavy ion bombardment. The formic acid half-life can be given by \citealt{barr2011}:

\begin{equation}
\label{eq:half}
\tau_{1/2} = \frac{ln(2)}{\sum_k \int\Phi_k(E)\sigma_{d,k}(E)dE}
\end{equation}
were \textit{E} is the kinetic energy of the projectile, $\Phi_k(E)$ is the estimated flux of cosmic rays between $E$ and $E + dE$, and $\sigma_{(d,k)}$ is the destruction cross section of the molecule $k$ (formic acid).

\begin{figure}
\centering
\resizebox{\hsize}{!}{\includegraphics{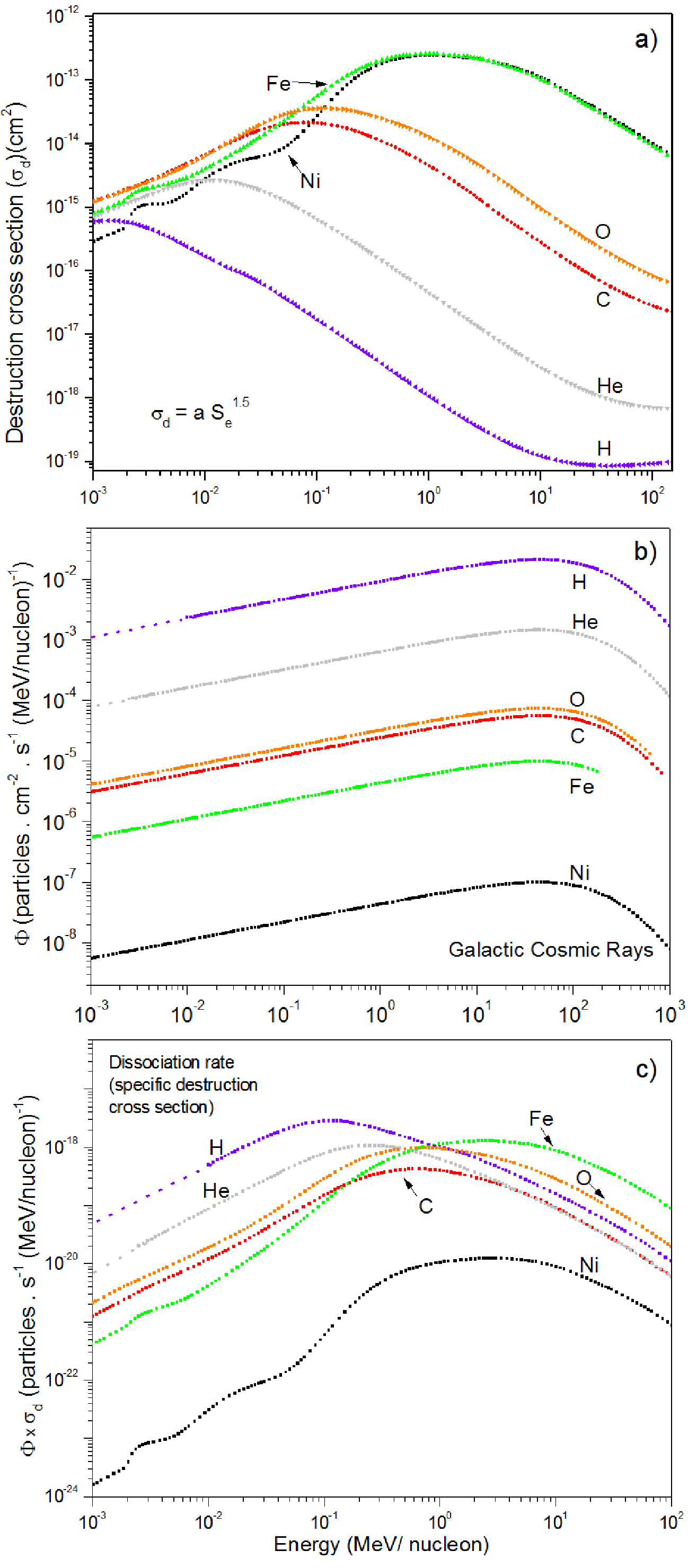}}
\caption{Energy dependence of main (H, He, C, O, Fe, and Ni) galactic cosmic ray constituents relative to (a) HCOOH destruction cross section $\sigma_{d,k}$; (b) flux density $\Phi_k$; and (c) dissociation rate defined as the product 	$\Phi_k~\sigma_{d,k}$}
\label{fig:half_life}
\end{figure}
The procedure to solve Eq.~\ref{eq:half} is described graphically in Figure~\ref{fig:half_life}. Fig.~\ref{fig:half_life}a) shows the dependence of the destruction cross section $\sigma_{d,k}(E)$ in function of the ion energy for six cosmic rays constituents: H, He, C, O, Fe, and Ni ions. Fig.~\ref{fig:half_life}{b) shows the energy distribution and flux density $\Phi_k(E)$ of the cosmic rays. The estimated cosmic rays flux is adapted from \cite{shen2004}. The product $\Phi_k(E)~\sigma_{d,k}(E)$ is in Fig.~\ref{fig:half_life}c). The integral $\int\Phi_k(E)$ is performed over the cosmic rays energy range of 10 KeV to 10 GeV. The quotient between ln(2) and the sum of $\int\Phi_k(E)\sigma_{d,k}(E)dE$ for all the selected cosmic ray species provides the solution for Equation~\ref{eq:half}.

The calculated half-life of the formic acid molecule in a H$_2$O:HCOOH ice in interstellar medium, due to six species of cosmic rays (H, He, C, O, Fe, and Ni), was approximately $8.3 \times 10^7 years$, which is about one order of magnitude less than the results obtained by \cite{andr2013} for pure formic acid ice. The explanation for this difference may lie in the fact that the water molecules, rather than be a shield against radiation, are also ionized by the cosmic rays, generating more radicals, which accelerates the formic acid destruction apparently by a factor of $\backsim3.5$.

\section{Conclusions} 

A formic acid-containing ice, at 15 K, irradiated by 46 MeV $^{58}$Ni$^{11+}$ ions, analogs to galactic cosmic rays, in an ultra-high vacuum regime, was studied in this work in order to understand the destruction of this important pre-biotic specie precursor, and the formation of new species in terms of surface astrochemistry. The analysis was made by FTIR spectrometry, and the results show intense formation of CO and CO$_2$ molecules due to the destruction of the parent molecule. The results of this experiment were compared to those of \cite{andr2013}, providing better understanding of the influence of water in astrophysical ices during heavy ion bombardment. The results shows that the destruction cross section ($\sigma_d$) in an H2O:HCOOH ice is slightly higher compared to the pure formic acid ice. The variation in the column densities, as well as the half-life in interstellar medium in the two experiments, are compatible. As shown in the section 3.4, the heating of the sample after bombardment produces new species, mostly hydrocarbons.

To estimate the half-life of formic acid in the ISM, we have studied the dissociation cross-section of the C-O (1211 cm$^{-1}$) vibration mode. To predict the half-life of formic acid due to the major constituent ions detected in cosmic rays, we have considered that the destruction cross-section follows a power law as a function of the electronic stopping power, as seen in Andrade et al. (2013). For this particular ice, we have considered $\sigma_d$ $\thicksim~a~S_e^n$, were n = 1.5. The estimated half-life time of the formic acid in water ice due to cosmic rays is approximately 83 million years.

\section{Acknowledgements}

This work was supported by the French-Brazilian exchange programme CAPES-COFECUB. The Brazilian agencies CNPq (INEspaço) and FAPERJ also provided partial support. We would like to thank A. Domaracka, E. Balanzat, T. Been, F. Levesque, T. Madi, I. Monnet, Y. Ngono-Ravache and J.M. Ramillon for their invaluable support. The author would like to thank to FAPESP agency for the financial support.

\bibliography{ref}
\bibliographystyle{mn2e}

\label{lastpage}

\end{document}